\documentclass[12pt]{article}
\usepackage[utf8]{inputenc}
\usepackage{rotating}
\usepackage{enumerate}
\usepackage{mathtools}
\usepackage{todonotes}

\usepackage{url}
\usepackage{amsfonts}
\usepackage{amsmath,amssymb,color}
\usepackage{graphicx}
\usepackage{epstopdf}
\usepackage{enumitem}
\usepackage{multirow}
\usepackage{natbib}

\usepackage{verbatim}
\usepackage{float}
\usepackage[toc,page]{appendix}
\usepackage{setspace}
\usepackage{adjustbox}
\usepackage{rotating}
\usepackage{subcaption}
\usepackage{amsthm}
\usepackage{booktabs,dcolumn,caption}

\usepackage{color}
\usepackage{tikz}
\usepackage{multirow}
\usepackage{graphicx}
\graphicspath{{figures/}}
\usepackage{adjustbox}
\usepackage{float}
\usepackage{footnote}
\usepackage{array}
\usepackage{longtable}
\usepackage{threeparttablex}
\usepackage{booktabs,dcolumn,caption}
\usepackage{amsmath,amssymb,color}
\usepackage{fnbreak}
\usetikzlibrary{arrows.meta,shapes.multipart,positioning}
\usepackage{mathtools}
\usepackage{arydshln}

\usepackage{mathabx}
\usepackage{array}
\usepackage{longtable}
\usepackage{threeparttablex}

\usepackage{chngcntr}
\usepackage{pdflscape}

\usepackage{tikz} 
\usepackage{algorithm}
\usepackage{algpseudocode}

\newcolumntype{L}[1]{>{\raggedright\let\newline\\\arraybackslash\hspace{0pt}}p{#1}}
\newcolumntype{C}[1]{>{\centering\let\newline\\\arraybackslash\hspace{0pt}}p{#1}}
\newcolumntype{R}[1]{>{\raggedleft\let\newline\\\arraybackslash\hspace{0pt}}p{#1}}

\newcommand{\bA}{\bm{A}}
\newcommand{\bB}{\bm{B}}
\newcommand{\bC}{\bm{C}}

\newcommand{\bE}{\bm{E}}

\newcommand{\bI}{\bm{I}}

\newcommand{\bQ}{\bm{Q}}

\newcommand{\bgamma}{\bm{\gamma}}

\newcommand{\indep}{\perp \!\!\! \perp}

\newcommand{\argmax}{\operatornamewithlimits{arg\,max}}

\newtheorem{theorem}{Theorem}

\newtheorem{assumption}{Assumption}

\newtheorem{condition}{Condition}

\newtheorem{proposition}{Proposition}

\usepackage[english]{babel}
\usepackage{amsthm}
\usepackage{bm}

\usepackage{authblk}
\usepackage[colorlinks=true,allcolors={[rgb]{0,0,0.55}}]{hyperref}
\title{\vspace{-1.2cm}\large Mental Health Support Hotlines as Capacity-Limited Services: Joint Modeling of Demand, Assessment, and Unassessed Calls}

\author[$\dagger$]{\small Siliang Zhang\thanks{The two authors contributed equally to this work.}}
\author[$\ddagger$]{\small Jing Ouyang$^{\ast}$}
\affil[$\dagger$]{\small Key Laboratory of Advanced Theory and Application in Statistics and Data Science-MOE, School of Statistics, East China Normal University}
\affil[$\ddagger$]{\small Faculty of Business and Economics, The University of Hong Kong}
\date{}

\begin{document}

\maketitle

\begin{abstract}

Mental health support hotlines log help-seeking call attempts in real time, yet limited service capacity allows only a small proportion of calls to be answered and assessed. The time of every attempt is recorded, whereas issue types and marks such as caller demographics and crisis severity are observed only for assessed calls. 
Caller anonymity also prevents linking repeated attempts to the same individual.
Nevertheless, service planning and crisis monitoring require estimating unassessed demand, recovering temporal patterns in call content, and quantifying how much the partially observed marks improve these estimates. We develop a Joint Marked Dynamic Factor Model (JM-DFM) in which a shared low-dimensional latent state drives attempt intensity, issue composition, and mixed-type mark distributions. We establish identifiability, prove consistency of the proposed sieve estimator, and show that incorporating marks improves the asymptotic estimation precision. 
The proposed framework is then applied to records of $68{,}371$ call attempts to China's national 12356 mental health support hotline. We estimate that about $35$ unassessed attempts per day involve suicidal ideation, representing a higher proportion than among assessed calls. Incorporating the marks narrows the latent-state uncertainty bands by 18\% to $49\%$, most during periods with few issue-labeled calls.

\bigskip
{\raggedright
\par}
\end{abstract}

\section{Introduction}

Timely access to mental health care is critical for people in psychological distress \citep{tadmon2023differential}, and providing such support is {a priority} for public mental health hotline services. 
To allocate limited hotline resources, government agencies and service providers rely on information about when calls arrive and the concerns and clinical risks these calls involve. 
Mental health support hotlines record help-seeking in real time \citep{brulhart2021mental} and can inform these resource allocation decisions. However, assessment records may include only part of the concerns and clinical risks associated with incoming call attempts. 
Nonetheless, service planning and crisis monitoring require 
characterizing the temporal patterns of demand, identifying the issue and risk profiles that remain unassessed, and evaluating how expanded access would alter the volume and composition of assessed attempts.


We analyze administrative records from the national 12356 mental health support hotline in China \citep{chinadaily2025hotline}, hereafter the 12356 hotline.
The data consist of a complete log of incoming call times, whereas structured assessment records are available only for calls that are answered and assessed. The assessment records include issue labels together with mixed-type clinical, demographic, and call characteristics, which we refer to as marks. 
When assessment probabilities and call content both vary over time, the pooled assessed sample can misrepresent {the issue and risk composition of incoming demand}.
Information on issue types and caller characteristics from assessed calls can be used to characterize unassessed attempts and incoming demand as a whole only after this selection is accounted for.
Callers are anonymous, so calls cannot be linked to the same person, and standard subject-level longitudinal or recurrent-event methods are not directly applicable. These features motivate a system-level analysis that takes the hotline as the unit of analysis and the incoming call stream as aggregate demand.

Existing work provides useful tools for modeling recurrent events observed from individuals.
Classical recurrent-event methods use counting-process, rate, and frailty models
\citep{andersen1982cox,cook2007statistical,lin2000semiparametric,lin2001semiparametric,liu2004shared},
with extensions for markers measured at recurrent events \citep{cai2010semiparametric}. Joint
longitudinal--event models link longitudinal measurements and event processes through shared
latent effects \citep{wulfsohn1997joint,henderson2000joint,kim2012joint}.  However, these individual-level approaches rely on repeated events or measurements being linked to the same individual over time and are therefore not directly applicable to the hotline records, where callers are anonymous.

A related literature studies service demand and temporal event processes. Applied work on crisis
hotlines, emergency-call systems, and call centers has examined temporal and geographic
variation in demand and forecast arrival rates
\citep{weinberg2007bayesian,matteson2011forecasting,zhou2015spatio,marco2017spatio,ehrlich2023trends}.
These studies characterize patterns in recorded demand, but generally do not focus on recovering the issue and risk
composition of attempts that receive no assessment. Marked point-process models extend the
analysis to event occurrences and associated labels or attributes
\citep{coxisham1980point,daley2003introduction}, with self-exciting and neural formulations
allowing dependence on event history
\citep{hawkes1971spectra,ogata1988statistical,mohler2011self,du2016recurrent,reinhart2018review,shchur2021neural}.
However, modeling event--mark dependence alone does not account for the fact that call content is observed only for a proportion of assessed calls. In parallel, dynamic and generalized factor
models use low-dimensional latent structure to capture common variation in high-dimensional
temporal and non-Gaussian data
\citep{forni2000generalized,stock2002forecasting,bai2003inferential,bartholomew2011latent,wang2022maximum,chenli2022determining}.
Recent work introduces dynamic factor structure into recurrent-event and multivariate counting
processes \citep{chen2025dynamic,chen2025dynamicfactor}, while joint latent space models show
how network data and node variables can jointly improve estimation of shared latent positions
\citep{zhang2022joint}. However, these approaches are primarily developed for settings in which all data components are recorded for every observational unit. In the hotline records, the complete stream of anonymous attempts captures system-level demand, whereas issue and risk information is only partially observed. 
 This observation pattern creates the inferential challenge of estimating the shared temporal structure and recovering the issue and risk composition of unassessed attempts.

Motivated by these challenges, we develop a Joint Marked Dynamic Factor Model (JM-DFM) that integrates the complete call-attempt stream with partially observed assessment records.
The model specifies attempt intensity, issue composition, and mixed-type mark distributions as functions of a shared low-dimensional latent trajectory with smooth long-term and periodic within-day components.
Factor trajectories describe the evolution of common temporal patterns, while loadings characterize their associations with call volume, issue types, and caller characteristics.
Separate assessment and recording mechanisms describe whether an attempt reaches assessment and which issue labels and marks are recorded.

The statistical contributions are threefold.  
First, we formulate an observed-data likelihood that combines all attempt times with the available issue labels and marks, summing over unrecorded labels and integrating out missing marks.
Under ignorability and positivity conditions, this formulation supports estimation of {the issue and risk composition of unassessed demand}. 
Combined with the assessment model, it yields estimands for unassessed demand and additional assessments under alternative access profiles at fixed demand and content distributions.
Second, we develop a B-spline sieve method for jointly estimating the long-term and within-day latent factors. We establish conditions for identifiability of the model and prove that the sieve estimator consistently recovers these latent trajectories, even when issue labels and marks are only partially observed.
Third, we characterize the information supplied by recorded marks beyond attempt times and observed issue labels. A local one-step analysis establishes conditions under which incorporating informative marks reduces the pointwise asymptotic mean squared error of latent-state estimation.

Records of 68,371 attempts at one site of the 12356 hotline show a mismatch between the timing of demand and access to assessment. Attempts are concentrated in evening and overnight hours, when assessment probabilities are lowest.
Under the observation assumptions, the model estimates a higher proportion involving suicidal ideation among unassessed attempts than among assessed calls, corresponding to about 35 unassessed attempts per day.
Holding demand and content distributions fixed, raising evening assessment probability to the daytime median yields an estimated 77 additional assessments per day, including seven involving suicidal ideation.
The fitted factors summarize long-term and within-day variation in call content, and incorporating marks narrows latent-state uncertainty most during periods with few issue-labeled calls.

The remainder of the paper is organized as follows. Section~\ref{sec:data} describes the hotline records and observation mechanism and formalizes the three substantive and statistical questions. Section~\ref{sec:method} develops the JM-DFM, the observed-data likelihood, and the estimation and inference procedure. Section~\ref{sec:theory} establishes identifiability, consistency, and the asymptotic precision gained from marks. Section~\ref{sec:analysis} applies the model to the hotline records, and Section~\ref{sec:discussion} concludes with limitations and extensions. Proofs, simulation studies, and implementation details are provided in the Supplementary Material.

\section{Data and Scientific Questions}\label{sec:data}
 
\subsection{The 12356 Hotline as a Capacity-Limited Service}
\label{sec:data:background}

\begin{figure}[t]
\centering
\begin{tikzpicture}[
  font=\small,
  box/.style={draw, rectangle, align=center, inner sep=4pt},
  obs/.style={box, fill=black!8},
  hid/.style={box, densely dashed},
  mod/.style={box, fill=black!16},
  lab/.style={font=\footnotesize, inner sep=1pt, text height=1.5ex, text depth=0pt},
  >=Stealth]
\node[obs, text width=18em] (attempts) at (0,0)
  {\textbf{Incoming call attempts}\\[1pt]
   start time recorded for every attempt};
\node[box] (observe) at (0,-1.5)
  {reaches structured assessment?};
\node[obs, text width=16em] (assessed) at (-3.8,-3.0)
  {\textbf{Assessed calls}\\[1pt]
   issue type and marks recorded};
\node[hid, text width=16em] (hidden) at (3.8,-3.0)
  {\textbf{Unassessed attempts}\\[1pt]
   issue type and marks unobserved};
\node[mod, text width=7.6cm+16em] (model) at (0,-5.1)
  {\textbf{JM-DFM}\\[1pt]
   incoming demand, assessment probability, issue composition, mark distributions\\[1pt]
   shared latent temporal factors and unassessed demand};
\draw[->] (attempts) -- node[lab, right=3pt] {limited service capacity} (observe);
\draw[->] (observe) -- node[lab, pos=0.6, above left=0pt and 1pt] {yes} (assessed);
\draw[->] (observe) -- node[lab, pos=0.6, above right=0pt and 1pt] {no} (hidden);
\draw[->] (assessed) -- (model);
\draw[->] (hidden) -- (model);
\end{tikzpicture}
\caption{Observation structure of the 12356 hotline records. Attempt times are recorded for every call, while issue labels and {marks} are available only for assessed calls. {Assessed calls and unassessed attempts both contribute to the likelihood of the JM-DFM in Section~\ref{sec:method}.}}
\label{fig:layers}
\end{figure}
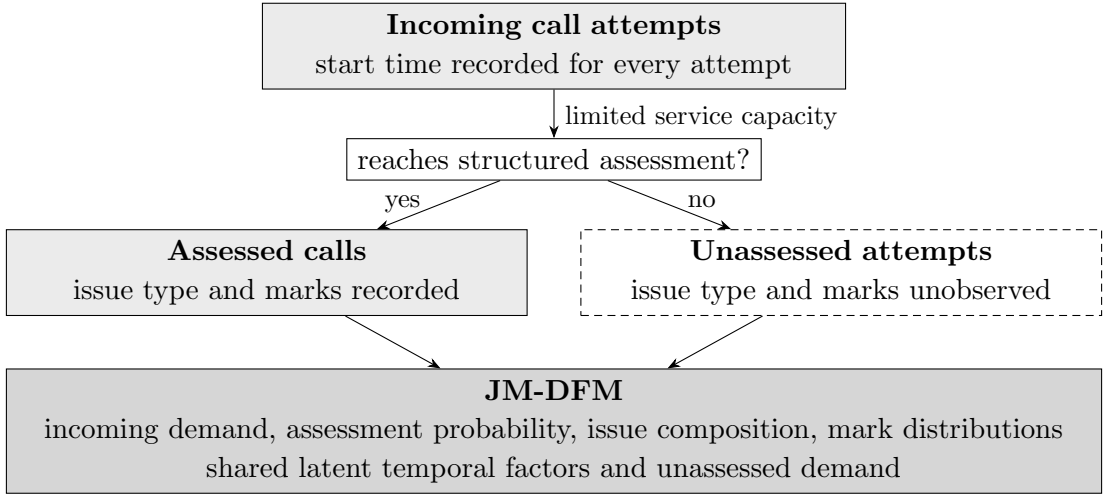

Mental health support hotlines provide critical, immediate support to individuals in distress and operate at a national scale in many countries. In China, the 12356 hotline was launched on May 1, 2025~\citep{chinadaily2025hotline}. We analyze call records from one site of this hotline across the 163-day period from June~21 to November~30, 2025, capturing its initial months of nationwide operation. To our knowledge, no statistical analysis of its records has been published.
The data provide two complementary views of this service. First, the site's telephone system recorded the start time of all $68{,}371$ incoming call attempts. Second, for the $14{,}676$ attempts that reached assessment, an operator completed a structured record giving the issue label and the clinical, demographic, and call-level characteristics of the call. An unassessed attempt therefore has a known time but unobserved content. Callers are anonymous and repeat attempts by the same person are not identified. Figure~\ref{fig:layers} summarizes this observation structure.

\begin{figure}[t]
\centering
\includegraphics[width=0.55\linewidth]{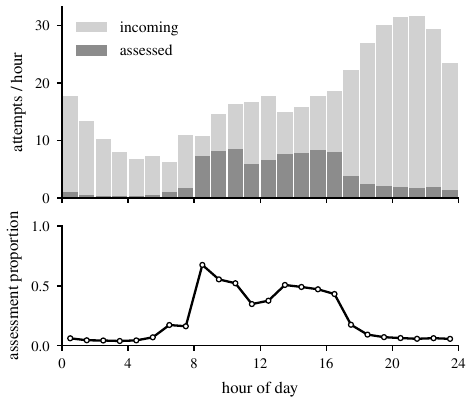}
\caption{Demand and assessment by hour of day. Mean daily counts of incoming and assessed attempts by hour (top) and the assessment proportion (bottom) over the $163$-day study window.}
\label{fig:access}
\end{figure}

\label{sec:data:access}%
Figure~\ref{fig:access} shows the proportion of attempts assessed by hour of day. About two thirds of attempts arrive in the evening or overnight, when this proportion falls far below its daytime level, {so issue labels and marks are observed for the smallest proportion of attempts when demand is highest.}
{The assessment proportion also varied over the study period, rising from $17.9\%$ in September to $28.1\%$ in November.}
{The issue composition and caller characteristics observed in the assessed records therefore reflect selection that varies over time.}
This temporal mismatch between demand and assessment access {motivates modeling call volume, issue composition, and the assessment probability jointly over time}.

\label{sec:data:structure}%

Each assessed call is characterized by a categorical issue type and a set of marks. Operators initially classify calls into one of 14 issue categories, with romance, marriage, and family problems the most common, followed by special calls such as silent or prank calls. Excluding two rare categories from content estimation leaves 12 issue types across 14{,}172 issue-labeled calls (Table~S4 of the Supplementary Material). In addition, operators record 11 candidate mark variables covering caller demographics, call characteristics, dominant emotion, and crisis severity measured on a five-level ordinal scale (ranging from no suicidal ideation to an active attempt). These marks capture clinical heterogeneity, distinguishing routine inquiries from acute crises within the same issue category. As summarized in Table~\ref{tab:marks-summary}, the marks vary in measurement scale and completeness. Our model retains ten of these variables (four of which are nominal), omitting the binary crisis flag because its information is fully subsumed by the ordinal crisis-severity mark.

\begin{table}[t]
\centering
\caption{Measurement scales, distribution families, and missingness of the marks among $14{,}676$ assessed calls. Operator-recorded ``unknown'' (not ascertained) responses are treated as missing.}
\label{tab:marks-summary}
\small
\begin{tabular}{llr}
\toprule
Mark variable & Scale (family) & Missing (\%) \\
\midrule
Gender         & Binary (Bernoulli) & 0.3 \\
Age            & Continuous (Gaussian) & 0 \\
Call duration  & Continuous (Gaussian, log scale) & 0 \\
Crisis flag    & Binary (Bernoulli) & 0 \\
Crisis severity & Ordinal, 5 levels (cumulative logit) & 0 \\
Calling on own behalf & Binary (Bernoulli) & 3.3 \\
Dominant emotion & Nominal, 8 levels (multinomial) & 26.2 \\
Education      & Ordinal, 6 levels (cumulative logit) & 42.8 \\
Marital status & Nominal, 4 levels (multinomial) & 20.7 \\
Employment     & Nominal, 22 levels (multinomial) & 37.5 \\
Living arrangement & Nominal, 3 levels (multinomial) & 19.4 \\
\bottomrule
\end{tabular}
\end{table}


\subsection{Scientific Questions and Target Estimands}\label{sec:data:questions}

{The observation structure motivates three scientific questions.} 

{
\begin{enumerate}[label=Q\arabic*.,ref=Q\arabic*,leftmargin=3em]

\item \emph{Unassessed demand.}\;
How many attempts go unassessed over time, and what issue and risk content do they involve?
We target the expected number of unassessed attempts in a time window, their {issue and risk composition}, and the additional assessments that higher assessment probabilities would yield at fixed demand and content distributions.  
\label{q:unmet}

\item \emph{Temporal structure in call content.}\;
What common temporal patterns underlie changes in issue composition and caller characteristics?
We target a low-dimensional system-level latent state with separate long-term and within-day components, interpreted through its loadings on the issue types and marks.\label{q:regimes}

\item \emph{Information contributed by the marks.}\;
How much do the recorded caller and call characteristics improve estimation of the latent temporal patterns?
We target the reduction in pointwise uncertainty of the estimated latent state relative to an events-only estimator that uses attempt times and recorded issue labels alone, particularly in periods with few issue-labeled calls.
\label{q:payoff}

\end{enumerate}}
To address these scientific questions, we next develop a joint modeling framework with a shared latent state. The latent state captures the common long-term and within-day variation in call volume, issue composition, and caller characteristics. Each recorded mark provides an additional measurement of this state. When assessment is independent of call content given the attempt time, the assessed calls also inform the content of unassessed attempts. 

\section{Joint Marked Dynamic Factor Model (JM-DFM)}\label{sec:method}

\subsection{Model Setup}
\label{sec:method:layers}\label{sec:method:shared}

Let $0<s_1<\cdots<s_N\le\tau$ denote the times of incoming call attempts on $[0,\tau]$.
For each call attempt $i$, whether assessed or unassessed, let
$\mathbf Y_i=(J_i,\mathbf W_i)$ denote its complete content, where $J_i\in\{1,\ldots,J\}$ is the
issue type and $\mathbf W_i=(W_{i1},\ldots,W_{iP})$
collects caller and call characteristics, referred to as marks.
We represent the system-level latent state at time $t$ by
$\mathbf Z(t)=\{\mathbf G(t)^\top,\mathbf H(u(t))^\top\}^\top\in\mathbb R^r$, where
$r=r_d+r_h$, $\mathbf G:[0,\tau]\to\mathbb R^{r_d}$ describes how the underlying system state evolves over the study period, $\mathbf H:[0,24)\to\mathbb R^{r_h}$ captures a periodic within-day pattern, and $u(t)=24(t\bmod 1)$ is the hour of day, with $t$ measured in days.

The attempt intensity, issue composition, and mark distributions are jointly linked via the shared latent state $\mathbf Z(t)$. First, let $N(t)$ be the
cumulative number of attempts by time $t$, and let $N=N(\tau)$. We model the call attempt as a Poisson process,
\begin{equation}\label{eq:demand}
N(dt)\sim\mathrm{Poisson}\{\lambda(t)\,dt\}, 
\qquad
\log\lambda(t)=\nu_0(t)+\boldsymbol\gamma^\top\mathbf Z(t),
\end{equation}
where the number of attempts in a short interval {$dt$} around time
$t$ follows a Poisson distribution {with mean $\lambda(t)\,dt$}. The function $\nu_0(t)$ is a low-dimensional baseline. 
The $\boldsymbol\gamma\in\mathbb R^r$ is the loading vector relating the shared latent state to the call-attempt intensity.

Second, for call attempt $i$, the issue type $J_i$ follows a categorical distribution,
\begin{equation}\label{eq:composition}
J_i\mid(s_i=t)\sim\mathrm{Categorical}\{q_1(t),\ldots,q_J(t)\},
\qquad
q_j(t)=\frac{\exp\{\mu_j+\boldsymbol\ell_j^\top\mathbf Z(t)\}}
{\sum_{k=1}^J\exp\{\mu_k+\boldsymbol\ell_k^\top\mathbf Z(t)\}},
\end{equation}
where $q_j(t)=\Pr(J_i=j\mid s_i=t)$ {is the probability that an attempt at time $t$ has issue type $j$. The vector $\{q_j(t)\}_{j=1}^J$ is the distribution of the issue type among attempts at time $t$, referred to as the issue composition.} Here $\mu_j$ is the intercept of issue type $j$ and
$\boldsymbol\ell_j=(\mathbf a_j^\top,\mathbf b_j^\top)^\top$ stacks its
long-term loadings $\mathbf a_j\in\mathbb R^{r_d}$ and within-day loadings
$\mathbf b_j\in\mathbb R^{r_h}$ so that $\boldsymbol\ell_j^\top\mathbf Z(t)
=
\mathbf a_j^\top\mathbf G(t)
+
\mathbf b_j^\top\mathbf H\{u(t)\}.$

Third, the marks are of mixed types.
We therefore model them using a generalized factor model in which $P$ marks are assumed conditionally
independent, with
\begin{equation}\label{eq:mark-model}
W_{ip}\mid(s_i=t,J_i=j)\sim
f_p\{\cdot\,;\delta_{pj}+\mathbf m_p^\top\mathbf Z(t)\},
\qquad p=1,\ldots,P.
\end{equation}
Here, $f_p(\cdot\,;\cdot)$ denotes the density or probability mass function
for mark $p$, whose distribution depends on its linear predictor $\delta_{pj}+\mathbf m_p^\top\mathbf Z(t)$. 
{In the application, $f_p$ is a Bernoulli, Gaussian, cumulative logit, or multinomial logit model according to whether mark $p$ is binary, continuous, ordinal, or nominal. Nuisance parameters such as Gaussian variances and ordinal thresholds are suppressed in the notation, and Section~S2.1 of the Supplementary Material gives the full specifications.}
The intercept $\delta_{pj}$ is issue specific, and $\mathbf m_p=(\mathbf c_p^\top,\mathbf e_p^\top)^\top$ stacks the
long-term loadings $\mathbf c_p\in\mathbb R^{r_d}$ and within-day
loadings $\mathbf e_p\in\mathbb R^{r_h}$. 

Together, \eqref{eq:demand}--\eqref{eq:mark-model} specify a marked Poisson process
with ground intensity $\lambda(t)$ whose issue type and marks at an
attempt time $t$ are drawn from \eqref{eq:composition} and
\eqref{eq:mark-model}.
Collect all model parameters as
$\Theta=(\nu_0,\boldsymbol\gamma,\boldsymbol\mu,A,B,
\boldsymbol\delta,C,E,\mathbf G,\mathbf H)$, with
intercepts $\boldsymbol\mu=(\mu_1,\ldots,\mu_J)^\top$ and
$\boldsymbol\delta=(\delta_{pj})_{p\le P,\,j\le J}$, and
loading matrices $A=(\mathbf a_1,\ldots,\mathbf a_J)^\top$,
$B=(\mathbf b_1,\ldots,\mathbf b_J)^\top$,
$C=(\mathbf c_1,\ldots,\mathbf c_P)^\top$, and
$E=(\mathbf e_1,\ldots,\mathbf e_P)^\top$. We
write $\lambda_\Theta(t)$ when the dependence on $\Theta$ is relevant.


\subsection{Missingness Mechanism and Observed-Data Likelihood}
\label{sec:method:decomposition}\label{sec:method:likelihood}
Missing data arise through two mechanisms. First, a call attempt may not reach the structured assessment stage. Second, even among assessed calls, only a subset of the issue label and marks may be recorded. {For an unassessed attempt, $\mathbf Y_i$ denotes the content that would have been recorded had the attempt been assessed. 

Let $R_i=1$ indicate that call attempt $i$ reaches structured assessment, and
let $\mathbf x_i$ denote operational variables observed for every attempt.
For an assessed call, let $\Omega_i\subseteq\{0,1,\ldots,P\}$ denote the index
set of recorded elements of $\mathbf Y_i$, where element $0$ is the issue
label and element $p\ge1$ is mark $W_{ip}$, and set
$\Omega_i=\varnothing$ when $R_i=0$. For $a\subseteq\{0,1,\ldots,P\}$,
write $\mathbf Y_{i,a}$ for the elements of $\mathbf Y_i$ in $a$. The
observed data are denoted as
$\mathcal O=\{s_i,\mathbf x_i,R_i,\Omega_i,\mathbf Y_{i,\Omega_i}\}_{i=1}^N$.


We assume that whether a call reaches assessment depends only on time and the
observed operational variables,
\begin{equation}\label{eq:access-model}
R_i\mid(s_i,\mathbf x_i,\mathbf Y_i,\mathbf Z)
\sim
\mathrm{Bernoulli}\{\pi_R(s_i,\mathbf x_i;\boldsymbol\phi_R)\},
\end{equation}
where the assessment probability $\pi_R(s_i,\mathbf x_i;\boldsymbol\phi_R)$ is modeled via logistic regression with parameter vector $\boldsymbol\phi_R$. 
Specification \eqref{eq:access-model} implies the conditional independence $R_i\indep\mathbf Y_i\mid(s_i,\mathbf x_i,\mathbf Z)$. Furthermore, we assume that call content is independent of operational variables given time, $\mathbf Y_i\indep\mathbf x_i\mid(s_i,\mathbf Z)$. Together, these assumptions ensure that the content models \eqref{eq:composition} and \eqref{eq:mark-model}, which condition on time alone, apply identically to both assessed and unassessed attempts.

Among assessed calls, we assume that the set of recorded variables
depends on time and the observed operational variables, independently of
the issue type and mark values,
\begin{equation}\label{eq:record-model}
\Pr(\Omega_i=a\mid R_i=1,s_i,\mathbf x_i,\mathbf Y_i,\mathbf Z)
=\pi_\Omega(a\mid s_i,\mathbf x_i;\boldsymbol\phi_\Omega),
\qquad a\subseteq\{0,1,\ldots,P\}.
\end{equation}
This assumption can be relaxed to allow the probability of a recording pattern $a$ to depend additionally on the recorded elements $\mathbf Y_{i,a}$, corresponding to a missing-at-random mechanism \citep{rubin1976inference}. The missingness parameters $\boldsymbol\phi=(\boldsymbol\phi_R,\boldsymbol\phi_\Omega)$ of $\pi_R$ and $\pi_\Omega$ are distinct from the model parameters $\Theta$.

If every attempt were assessed and fully recorded,
equations \eqref{eq:demand}--\eqref{eq:mark-model} would yield the complete-data
log likelihood
\begin{equation}\label{eq:complete-lik}
\ell_{\mathrm c}(\Theta)
=-\int_0^\tau\lambda_\Theta(t)\,dt
+\sum_{i=1}^N\Big[\log\lambda_\Theta(s_i)+\log q_{j_i}(s_i)
+\sum_{p=1}^P\log f_p\{w_{ip};\delta_{pj_i}+\mathbf m_p^\top\mathbf Z(s_i)\}\Big].
\end{equation}
{Under \eqref{eq:access-model}, with independent access across attempts and $\pi_R$ depending on time, the assessed attempts of issue type $j$ form an independently thinned Poisson process with intensity $\lambda_\Theta(t)\pi_R(t)q_j(t)$.} 
Furthermore, under \eqref{eq:access-model} and \eqref{eq:record-model}, the assessment and recording mechanisms are ignorable for likelihood inference on $\Theta$, since they do not depend on the unobserved issue labels or marks and their parameters are distinct from $\Theta$ \citep{rubin1976inference,heitjan1991ignorability}. The observed-data likelihood therefore marginalizes over the unrecorded content.

Let $\mathcal J_i=\{j_i\}$ if the issue label is recorded and
$\mathcal J_i=\{1,\ldots,J\}$ otherwise. The likelihood contribution from the
recorded content of an attempt at time $t$ is
\begin{equation}\label{eq:content-mixture}
p_\Theta(\mathbf y_{i,\Omega_i}\mid t)
=
\sum_{j\in\mathcal J_i}q_j(t)
\prod_{p\in\Omega_i\setminus\{0\}}
f_p\{w_{ip};\delta_{pj}+\mathbf m_p^\top\mathbf Z(t)\}.
\end{equation}
If the issue label is unrecorded, the sum is over all possible issue types;
unrecorded marks contribute no factor. When no issue label or mark is
recorded, $p_\Theta(\mathbf y_{i,\varnothing}\mid t)=1$. Therefore,
\begin{equation}\label{eq:ign-lik}
\ell_{\mathrm{obs}}(\Theta)
=
-\int_0^\tau\lambda_\Theta(t)\,dt
+\sum_{i=1}^N
\left\{
\log\lambda_\Theta(s_i)
+\log p_\Theta(\mathbf y_{i,\Omega_i}\mid s_i)
\right\}.
\end{equation}

The assessment probability is estimated separately by penalized logistic
regression of $R_i$ on $(s_i,\mathbf x_i)$ over all $N$ attempts. In the
application $\pi_R$ depends on calendar time,
hour of day, and day-of-week indicators, all of which are functions of
$s_i$, so $\pi_R(t,\mathbf x)$ reduces to a function of time alone $\pi_R(t)$.

\subsection{Model Identifiability}
\label{sec:identifiability}

Identifiability of the JM-DFM requires additional care beyond standard
factor analysis~\citep{bai2003inferential,bai2012statistical}. In our
setting, the same {latent state $\mathbf Z$ is shared} across the call attempt intensity,
issue composition, and mark distributions through different sets of
loading parameters. Consequently, a change of coordinates in a latent
trajectory can be absorbed simultaneously by several loading matrices
without changing the observed-data distribution. In addition, the
long-term and within-day latent trajectories each admit their own location,
scale, and rotation indeterminacies.

Specifically, for any invertible matrices $\bQ_d$ and $\bQ_h$, the following
transformations
\begin{align*}
    \mathbf G(t)\mapsto \bQ_d^{-1}\mathbf G(t), \qquad
(\bA,\bC,\bgamma_d)\mapsto
(\bA\bQ_d,\bC\bQ_d,\bQ_d^\top\bgamma_d);\\
\mathbf H(u)\mapsto \bQ_h^{-1}\mathbf H(u), \qquad
(\bB,\bE,\bgamma_h)\mapsto
(\bB\bQ_h,\bE\bQ_h,\bQ_h^\top\bgamma_h),
\end{align*}
leave the corresponding distributions unchanged. Hence, without additional
restrictions, {the latent state and its loading parameters} are
identified only up to these transformations.

We impose the following normalization conditions to fix these indeterminacies.
Identifiability also uses the temporal separation and observation
conditions stated in Section~\ref{sec:theory}.

\begin{condition}[Identifiability]
\label{cond:identifiability}
The loading matrices $\bA$, $\bB$, $\bC$, and $\bE$ have full column rank,
{the intercepts satisfy} $\sum_j\mu_j=0$, {the issue loadings satisfy} $\sum_j\boldsymbol\ell_j=\mathbf0$, and the first threshold of each ordinal mark is fixed at zero.
Moreover,
\begin{equation}\label{eq:normalization}
\begin{aligned}
J^{-1}\bA^\top\bA&=\bI_{r_d},\qquad J^{-1}\bB^\top\bB=\bI_{r_h},\qquad
\int_0^\tau \mathbf G(t)\,dt=\mathbf0,\qquad
\int_0^{24}\mathbf H(u)\,du=\mathbf0,\\
P^{-1}\bC^\top\bC
&=\operatorname{diag}(\kappa_{d1},\ldots,\kappa_{dr_d}),
\qquad
\kappa_{d1}>\cdots>\kappa_{dr_d}>0,\\
P^{-1}\bE^\top\bE
&=\operatorname{diag}(\kappa_{h1},\ldots,\kappa_{hr_h}),
\qquad
\kappa_{h1}>\cdots>\kappa_{hr_h}>0.
\end{aligned}
\end{equation}
\end{condition}

The centering constraints separate the latent temporal variation from the
baseline terms, while the remaining normalizations fix the scales, rotations,
and ordering of the latent coordinates. Note that the remaining sign indeterminacy in factor analysis does not affect the interpretation of the model as the relationships between observed processes and latent states are unchanged. Following the usual convention, we assume that the estimated latent directions have the same signs as the corresponding true latent directions \citep{bai2012statistical}.

Let $\mathcal N
=
\{\Theta:\Theta\text{ satisfies Condition~\ref{cond:identifiability}}\}$ {denote the normalized parameter space}.
All estimation and inference below are carried out over $\mathcal N$.

\subsection{Estimation and Inference}
\label{sec:method:estimation}
\label{sec:estimation}

We estimate the model by maximizing the observed-data likelihood over the
normalized parameter space introduced in Section~\ref{sec:identifiability}.
To approximate the latent trajectories, we use cubic B-spline sieves,
\begin{equation}\label{eq:calendar-sieve-residualized}
\mathbf G(t)
=
\Gamma_d^\top\mathbf b_d(t),
\qquad
\mathbf H(u)
=
\Gamma_h^\top\mathbf b_h(u),
\end{equation}
where $\Gamma_d\in\mathbb R^{K_d\times r_d}$ and $\Gamma_h\in\mathbb R^{K_h\times r_h}$ are coefficient matrices. Here, $\mathbf b_d$ is constructed from a clamped cubic B-spline basis on $[0,\tau]$, and $\mathbf b_h$ from a periodic cubic B-spline basis on $[0,24)$. The functions in $\mathbf b_h$ are centered over the day, and those in $\mathbf b_d$ are orthogonalized in $L^2[0,\tau]$ with respect to $\mathbf b_h$ and the baseline functions in $\nu_0$. 
Let $\mathcal N_n$ denote the set of parameters in $\mathcal N$ whose trajectories have the form \eqref{eq:calendar-sieve-residualized}. The JM-DFM estimator is
\begin{equation}\label{eq:primary-estimator}
\widehat\Theta
\in
\argmax_{\Theta\in\mathcal N_n}
\ell_{\mathrm{obs}}(\Theta).
\end{equation}

In terms of computation,
{we maximize the observed-data likelihood via gradient ascent, centering the state and rescaling the loadings during the iterations, and resolve the remaining rotational indeterminacy afterwards so that the estimate satisfies Condition~\ref{cond:identifiability} (see Section~S2.1 of the Supplementary Material).}
For interpretation, we additionally report an equivalent representation
obtained by applying Varimax rotations \citep{kaiser1958varimax} to $A$ and
$B$, with the corresponding rotations applied to the latent states and
loadings. These rotations leave the fitted linear predictors unchanged but
need not preserve the ordered diagonal restrictions defining
$\mathcal N_n$. The factor dimensions $(r_d,r_h)$ are selected by
cross-validation.




We quantify uncertainty using a parametric bootstrap
\citep{efron1993introduction,davison1997bootstrap} that regenerates the
observed data from the fitted model. For $b=1,\ldots,M$, draw attempt times
$s_i^{(b)}$ from a Poisson process with intensity $\lambda_{\widehat\Theta}$,
assessment indicators $R_i^{(b)}$ from the fitted access probability, and
recording indicators from their empirical frequencies. For each assessed
call, draw
\begin{equation}\label{eq:bootstrap-draw}
\begin{aligned}
J_i^{(b)}&\sim\mathrm{Categorical}\{q_1(s_i^{(b)};\widehat\Theta),\ldots,
q_J(s_i^{(b)};\widehat\Theta)\},\\
W_{ip}^{(b)}\mid J_i^{(b)}=j
&\sim f_p\{\cdot\,;\widehat\delta_{pj}
+\widehat{\mathbf m}_p^\top\widehat{\mathbf Z}(s_i^{(b)})\},
\qquad p\in\Omega_i^{(b)}\setminus\{0\},
\end{aligned}
\end{equation}
and refit the access model and \eqref{eq:primary-estimator} on the
resulting data to obtain $\widehat\pi_R^{(b)}$ and $\widehat\Theta^{(b)}$.
Percentile intervals for a functional such as $q_j(t)$ or the estimands of
Section~\ref{sec:analysis:hidden} are formed from its $M$ replicate values.
The intervals therefore reflect {uncertainty in the fitted demand intensity,}
{access probability, issue composition, and mark distributions}. For the latent state and loadings, each
bootstrap estimate is first aligned with the reported estimate by an
orthogonal Procrustes rotation.

\section{Theoretical Results}
\label{sec:ident}
\label{sec:theory}

In this section, we establish the theoretical properties of the JM-DFM. We first establish identifiability guarantees and consistency of the joint sieve estimator. We further quantify the asymptotic precision gain from incorporating the marks.
Throughout the asymptotic analysis, the true parameter $\Theta^\star$ is
taken to belong to $\mathcal N$, and $\widehat\Theta$ denotes the corresponding
normalized estimator defined in \eqref{eq:primary-estimator}.

\subsection{Consistency of the Estimated Latent State}
\label{sec:ident:estimation}

We first establish consistency of the estimated latent state, showing that the proposed sieve estimator recovers the underlying temporal dynamics as the amount of observed event information increases. To formulate the results, let $\mathcal N_n$ denote the normalized parameter
space in \eqref{eq:primary-estimator} with spline dimensions
$(K_{d,n},K_{h,n})$.
Write
$
\widehat{\mathbf Z}(t)=
\big[\widehat{\mathbf G}(t)^\top,
\widehat{\mathbf H}\{u(t)\}^\top\big]^\top,
$ and $
\|\mathbf Z\|_{\mathcal Z}^2=
\tau^{-1}\int_0^\tau\|\mathbf G(t)\|_2^2dt+
24^{-1}\int_0^{24}\|\mathbf H(u)\|_2^2du.
$

We present the following assumptions that characterize the regularity requirements needed to control the spline approximation and to consistently recover the latent temporal dynamics.

\begin{assumption}
\label{ass:separation}
Let $\mathcal D$ be the finite-dimensional baseline space and
$\mathcal H$ the centered $24$-hour periodic space lifted to
$[0,\tau]$. The baseline functions are linearly independent, $\mathcal D\cap\mathcal H=\{0\}$, and $\mathcal G=(\mathcal D+\mathcal H)^\perp$.
Moreover
\[
\lambda_{\min}\!\left\{
  \tau^{-1}\int_0^\tau
  \mathbf G^\star(t)\mathbf G^\star(t)^\top\,dt
\right\}>0,
\qquad
\lambda_{\min}\!\left\{
  24^{-1}\int_0^{24}
  \mathbf H^\star(u)\mathbf H^\star(u)^\top\,du
\right\}>0.
\]
\end{assumption}

Assumption~\ref{ass:separation} separates the baseline, long-term, and within-day components and ensures that the specified latent states are nondegenerate. In particular, the positive-definiteness conditions require each latent direction to exhibit nonzero temporal variation.

\begin{assumption}
\label{ass:recording-support}
There exists $c>0$ such that $
\inf_{t,\mathbf x}\pi_R(t,\mathbf x)\ge c .$
For each mark $p$, and for the issue label itself when $p=0$,
there exists a recording pattern $a_p$ that records the issue label and
mark $p$ and satisfies $
\Pr\{\Omega_i=a_p\mid
R_i=1,s_i,\mathbf x_i,\mathbf Y_i,\mathbf Z\}
=
\rho_{a_p}(s_i,\mathbf x_i)\ge c .
$
\end{assumption}

Assumption~\ref{ass:recording-support} ensures that each issue or mark
component is observed with nonvanishing probability without
content-dependent selection. Other recording patterns may follow the missing at random mechanism in
\eqref{eq:record-model}.

\begin{assumption}
\label{ass:regularity}
The finite-dimensional parameters lie inside a compact space, and all linear predictors are uniformly bounded.
For each mark $p$, let
$\ell_p(w;\eta,\theta_p)$ denote its log density, which is identifiable in $(\eta,\theta_p)$, and
$\mathcal I_p(\eta,\theta_p)$ its Fisher information. Uniformly over the parameter set, $
cI\preceq
\mathcal I_p(\eta,\theta_p)
\preceq CI$
for some constants $0<c<C$, and its first
three derivatives have uniformly sub-exponential tails.

\end{assumption}

Assumption~\ref{ass:regularity} imposes boundedness on the finite-dimensional parameters and linear predictors and regularity conditions on the retained mark families. Closely related smoothness and curvature conditions are commonly used in generalized factor models and hold for a wide range of generalized frameworks \citep{wang2022maximum}. For each mark model, the information bounds ensure nondegenerate local curvature, while the derivative and tail conditions ensure uniform concentration of its score and Hessian.

\begin{assumption}
\label{ass:sieve}
For some integer $m\le4$, each coordinate of $\mathbf G^\star$ and
$\mathbf H^\star$ has $m$ bounded continuous derivatives. The long-term and
periodic cubic B-spline bases $\mathbf b_d$ and $\mathbf b_h$ have
quasi-uniform knots and dimensions $K_{d,n}\asymp K_{h,n}\asymp K_n$,
$K_n\to\infty$, $n^{-1}K_n^2\log n\to0$, where $n$ scales the attempt
intensity $n\lambda_{\Theta^\star}(t)$.
\end{assumption}

Assumption~\ref{ass:sieve} controls both the approximation error and the
complexity of the growing spline sieve. In particular, the smoothness
condition yields a spline approximation error of order $O(K_n^{-m})$,
while the growth condition on $K_n$ ensures that the sieve dimension
increases sufficiently slowly relative to the amount of information in
the data. 
We first establish identifiability of the JM-DFM under these conditions.

\begin{proposition}[Identifiability]\label{prop:identifiability-revised}
Suppose Assumptions~\ref{ass:separation}--\ref{ass:regularity} and Condition~\ref{cond:identifiability} hold. Let $(\widetilde\Theta,\widetilde{\boldsymbol\phi})$ and
$(\Theta^\star,\boldsymbol\phi^\star)$ satisfy Condition~1 and induce the
same distribution of the observed data.
Then, $\widetilde\Theta=\Theta^\star$.
In particular, $\widetilde{\mathbf Z}(t)=\mathbf Z^\star(t)$ for every
$t\in[0,\tau]$.
\end{proposition}

Proposition~\ref{prop:identifiability-revised} shows that, under the identifiability condition (Condition~\ref{cond:identifiability}), the joint model
parameters, and hence the shared latent trajectory, are uniquely determined
by the observed-data distribution.  

\begin{theorem}[Consistency of the sieve estimator]\label{thm:G}
Under Assumptions~\ref{ass:separation}--\ref{ass:sieve} and Condition~\ref{cond:identifiability}, for the estimator defined in \eqref{eq:primary-estimator}, we have
\[
\|\widehat{\mathbf Z}-\mathbf Z^\star\|_{\mathcal Z}^2
=O_p\!\left(K_n^{-2m}+\frac{K_n\log n}{n}\right)=o_p(1).
\]
In particular, if $K_n\asymp(n/\log n)^{1/(2m+1)}$, then
$\|\widehat{\mathbf Z}-\mathbf Z^\star\|_{\mathcal Z}^2
= O_p\{(\log n/n)^{2m/(2m+1)}\}$. 
\end{theorem}

Theorem~\ref{thm:G} establishes consistent recovery of the shared latent
trajectory from the joint observed-data likelihood despite partial
observation of the issue and mark information. The two terms in the
convergence rate correspond to spline approximation error and estimation
error, respectively. Balancing these terms yields the usual one-dimensional nonparametric rate for an \(m\)-smooth trajectory, up to a logarithmic factor \citep{stone1982optimal}.

\subsection{{Asymptotic} Precision Gained from Marks}
\label{sec:ident:info}

We next quantify the additional precision in estimating the latent state
contributed by the recorded marks.
To isolate the contribution of the marks, write \eqref{eq:ign-lik} as
$\ell_{\mathrm{obs}}=\ell_E+\ell_M$, where
\begin{equation}\label{eq:mark-lik}
\ell_M(\Theta)
=\sum_{i=1}^N\log
\frac{p_\Theta(\mathbf y_{i,\Omega_i}\mid s_i)}
{p_\Theta(\mathbf y_{i,\Omega_i\cap\{0\}}\mid s_i)}
\end{equation}
is, under content-independent recording, the conditional log likelihood of
the recorded marks given $\mathcal F_E$, the information generated by the
attempt times, recording patterns, operational covariates, and recorded
issue labels, and $\ell_E=\ell_{\mathrm{obs}}-\ell_M$ is the log
likelihood of the attempt times and recorded issue labels. For a mark weight
$\omega\ge0$, let
\begin{equation}\label{eq:weighted-lik}
\ell_\omega(\Theta)=\ell_E(\Theta)+\omega\,\ell_M(\Theta).
\end{equation}
Specifically, $\omega=0$ excludes the marks, whereas $\omega=1$ gives the full
observed-data likelihood. Intermediate values $0<\omega<1$ provide a
continuous way to quantify how incorporating the mark information changes
the precision of the estimated state.

Recall that the JM-DFM estimator in \eqref{eq:primary-estimator} estimates
the spline coefficient matrices in \eqref{eq:calendar-sieve-residualized}.
Let $\boldsymbol\beta$ collect these coefficients, so that
$\mathbf Z_{\boldsymbol\beta}(t)=\mathcal B_n(t)\boldsymbol\beta$.
Write $L_{E,n}=n^{-1}\ell_E$ and $L_{M,n}=n^{-1}\ell_M$ on the sieve.
Let $S_E,S_M$ and $H_E,H_M$ denote the corresponding state scores and
information matrices, with only the family-specific nuisance scores for
$\theta_p$ in Assumption~\ref{ass:regularity} projected out of $S_M$.
The finite-dimensional intercepts and loadings are held fixed in this
state-information comparison; score orthogonality under content-independent recording is
established in Lemma~S5 of the Supplementary Material.

We next start from the estimator using only the event--issue
information and update it in the direction contributed by the mark score.
Let $\widetilde{\boldsymbol\beta}_E$ denote the sieve estimator
based on $\ell_E$. We incorporate
the mark information through the one-step update
\begin{equation}
\label{eq:one-step-estimator}
\widetilde{\boldsymbol\beta}_{\omega,\mathrm{os}}
=
\widetilde{\boldsymbol\beta}_E
+
(\widehat H_E+\omega\widehat H_M)^{-1}
\omega\widehat S_M(\widetilde{\boldsymbol\beta}_E),
\qquad
\widetilde{\mathbf Z}_{\omega,\mathrm{os}}(t)
=
\mathcal B_n(t)\widetilde{\boldsymbol\beta}_{\omega,\mathrm{os}} .
\end{equation}

The one-step estimator in \eqref{eq:one-step-estimator} has first-order
coefficient covariance $n^{-1}V_{\omega,n}$, where
$
V_{\omega,n}
=
(H_E+\omega H_M)^{-1}
(H_E+\omega^2H_M)
(H_E+\omega H_M)^{-1}.
$
Since $\mathbf Z(t)=\mathcal B_n(t)\boldsymbol\beta$, its first-order pointwise covariance is
$n^{-1}\mathcal B_n(t)V_{\omega,n}\mathcal B_n(t)^\top$.
We therefore compare the precision under different mark weights through
the corresponding pointwise quadratic risk
\[
\operatorname{AMSE}_{\omega,n}(t)
=
\frac1n
\operatorname{tr}\!\left\{
\mathcal B_n(t)V_{\omega,n}\mathcal B_n(t)^\top
\right\}.
\]
The derivation of the one-step expansion and $V_{\omega,n}$ is given in
the Supplementary Material.


\begin{theorem}[AMSE gain from marks]
\label{thm:amse}

Suppose Assumptions~\ref{ass:separation}--\ref{ass:sieve} and
Condition~\ref{cond:identifiability} hold, and the recording mechanism
is content-independent as in \eqref{eq:record-model}. For each fixed $t\in(0,\tau)$ and
$0\le\omega_1<\omega_2\le1$, for all sufficiently large $n$, $V_{\omega_2,n}\prec V_{\omega_1,n}$, and $\operatorname{AMSE}_{\omega_2,n}(t)
<
\operatorname{AMSE}_{\omega_1,n}(t).$
In particular, for every fixed $\omega\in(0,1)$,
\[
V_{1,n}\prec V_{\omega,n}\prec V_{0,n},
\qquad
\operatorname{AMSE}_{1,n}(t)
<
\operatorname{AMSE}_{\omega,n}(t)
<
\operatorname{AMSE}_{0,n}(t).
\]
\end{theorem}

Theorem~\ref{thm:amse} shows that informative marks improve the
first-order precision of the estimated latent state {even when they are only partially observed}.
In practice, however, the mark models may be imperfectly specified,
particularly when richer features extracted from call audio or transcripts
are incorporated. The resulting precision gain may then be accompanied
by estimation bias, reflecting the familiar tradeoff between efficiency
under correct specification and robustness to misspecification~\citep{hausman1978specification}. To characterize this bias--variance tradeoff, we consider
a post-estimation interpolation between the events-only and joint estimators~\citep{green1991james}.


\begin{proposition}[Optimal interpolation weight]
\label{prop:optimal-interpolation}
Let $\widetilde{\mathbf Z}_0$ and $\widetilde{\mathbf Z}_1$ be the state estimators with $\omega=0$ and $\omega=1$, aligned as in Section~S3.4 of the Supplementary Material, and let $\widetilde{\mathbf Z}_\alpha=(1-\alpha)\widetilde{\mathbf Z}_0+\alpha\widetilde{\mathbf Z}_1$.
{The leading-order quadratic risk of $\widetilde{\mathbf Z}_\alpha(t)$ is minimized at the weight $\alpha_{\mathrm{opt}}(t)$ given by (S18) of the Supplementary Material.} Under correct specification and with negligible bias in spline approximation,
$\alpha_{\mathrm{opt}}(t)=1$. Under local misspecification confined to
the mark models, the optimal weight $\alpha_{\mathrm{opt}}(t) \in (0,1)$ when the
mark-induced bias is nonzero.
\end{proposition}
Proposition~\ref{prop:optimal-interpolation} complements
Theorem~\ref{thm:amse}. Under correct specification, the optimal
interpolation reduces to the joint estimator. Under local misspecification,
the optimal weight instead balances the precision gained from the marks
against the bias introduced through the mark model, placing less weight
on the joint estimator as the mark-induced bias increases.

\section{Analysis of the 12356 Hotline Records}\label{sec:analysis}

We now apply the JM-DFM to model the 12356 hotline records and address the three questions of Section~\ref{sec:data:questions} in turn. Simulation studies calibrated to the scale and recording pattern of these records (Section~S1 of the Supplementary Material) show accurate estimation with near-nominal interval coverage. Under the ignorability assumption \eqref{eq:access-model}, low assessment probabilities such as those in the evening and overnight still yield nearly unbiased estimates.

\subsection{Model Fitting and Selection}
\label{sec:analysis:preprocess}

The analysis combines all $68{,}371$ recorded attempt times with the available assessment data, including $J=12$ issue types recorded for $14{,}172$ calls and $P=10$ marks.
The spline bases in \eqref{eq:calendar-sieve-residualized} have $K_d=K_h=12$ functions. We select the numbers of factors by cross-validation over $(r_d,r_h)\in\{1,\ldots,4\}\times\{0,\ldots,3\}$, using the mean held-out content log likelihood per validation call. We form the folds by dividing the assessed calls that have an issue label into groups. For validation calls, we withhold the issue labels and marks from the fit but keep the attempt times in the demand likelihood. Cross-validation selects $(r_d,r_h)=(3,2)$. We then fit the JM-DFM based on the observed-data log likelihood \eqref{eq:weighted-lik} with $\omega=1$. Section~S2.5 of the Supplementary Material reports fits with other mark weights, and Section~S2.1 gives computational details and diagnostics for the choice of $(r_d,r_h)$.

The fitted demand intensity reproduces the observed day-of-week and within-day attempt patterns. Under the ignorability assumption, we estimate the access probability $\pi_R(t)$ from the assessment indicators in a separate step and use it to compute the estimands in \eqref{eq:estimands-main}. We quantify uncertainty using $200$ replicates of the parametric bootstrap in Section~\ref{sec:method:estimation}. Each replicate simulates new attempt times, assessment indicators, issue labels, and marks before refitting the access model and the JM-DFM. Every component of the estimated demand loading $\widehat{\boldsymbol\gamma}$ has a $95\%$ bootstrap interval that excludes zero, so attempt intensity is informative about each latent factor.

\subsection{Unassessed Attempts and Evening-Access Scenarios}
\label{sec:analysis:hidden}

The first question concerns unassessed demand (\ref{q:unmet}). Its estimands are functionals of the fitted demand intensity
$\lambda_\Theta(t)$, access curve $\pi_R(t)$, issue composition $\{q_j(t)\}_{j=1}^J$,
and mark model. For a time window $\mathcal I\subset[0,\tau]$, let
\begin{align}
D(\mathcal I)&=\int_{\mathcal I}\lambda_\Theta(t)\,dt,
&
U(\mathcal I)&=\int_{\mathcal I}\lambda_\Theta(t)\{1-\pi_R(t)\}\,dt,
\nonumber\\
U_{\mathcal R}(\mathcal I)&=\int_{\mathcal I}\lambda_\Theta(t)\{1-\pi_R(t)\}
\sum_{j=1}^J q_j(t)\,p_{\mathcal Rj}(t)\,dt,
&
\Delta^a(\mathcal I)&=\int_{\mathcal I}\lambda_\Theta(t)\{\pi_a(t)-\pi_R(t)\}\,dt,
\label{eq:estimands-main}
\end{align}
where $p_{\mathcal Rj}(t)=\Pr_\Theta\{\mathbf W_i\in\mathcal R\mid J_i=j,\mathbf Z(t)\}$
is the probability under the mark model \eqref{eq:mark-model} that a
type-$j$ call at state $\mathbf Z(t)$ has clinical profile $\mathcal R$.
Throughout, $\mathcal R$ is the suicidal-ideation profile, which covers any recorded crisis-severity level above no suicidal ideation, from ideation without a plan to an active attempt. Here
$D$ and $U$ are the expected numbers of incoming and unassessed attempts, and $U_{\mathcal R}$ is the expected number of unassessed attempts with profile $\mathcal R$. The quantity
$\Delta^a$ is a fixed-demand scenario that depends on a hypothetical
access curve $\pi_a(t)$. For instance, when $\pi_a$
is the fitted curve with the evening access probability raised to the
daytime median, $\Delta^a(\mathcal I)$ is the number of additional attempts
that the model implies would have been assessed over $\mathcal I$ at that
access level, with the attempt intensity and the content distribution held
at their fitted values. The form under covariate-dependent access is given
in Section~S2.7 of the Supplementary Material.

\begin{table}[t]
\centering
\caption{Estimated incoming and unassessed attempts \eqref{eq:estimands-main} by time window over the $163$-day study period, with $\mathcal R$ denoting the suicidal-ideation profile. The public-holiday period includes hours from each within-day window.}
\label{tab:hotspots}
\small
\setlength{\tabcolsep}{4pt}
\begin{tabular}{lrrrrr}
\toprule
Window & \shortstack[r]{Incoming\\attempts\\($D$)} & \shortstack[r]{Unassessed\\attempts\\($U$)} & \shortstack[r]{Unassessed\\proportion\\($U/D$)} & \shortstack[r]{Unassessed with\\suicidal ideation\\($U_{\mathcal R}$)} & \shortstack[r]{Ideation\\proportion\\($U_{\mathcal R}/U$)} \\
\midrule
Daytime (08--17) & $23{,}539$ & $12{,}700$ & $54.0\%$ & $1{,}027$ & $8.1\%$ \\
Evening (17--24) & $31{,}816$ & $29{,}169$ & $91.7\%$ & $2{,}711$ & $9.3\%$ \\
Overnight (00--08) & $13{,}016$ & $11{,}907$ & $91.5\%$ & $1{,}944$ & $16.3\%$ \\
After hours (17--08) & $44{,}832$ & $41{,}076$ & $91.6\%$ & $4{,}655$ & $11.3\%$ \\
\shortstack[l]{Major public-holiday\\period (8 days)} & $3{,}542$ & $2{,}714$ & $76.6\%$ & $275$ & $10.1\%$ \\
\midrule
Full study period & $68{,}371$ & $53{,}777$ & $78.7\%$ & $5{,}682$ & $10.6\%$ \\
\bottomrule
\end{tabular}
\end{table}

{The fitted intensity integrates to the $68{,}371$ recorded attempts, and Table~\ref{tab:hotspots} reports the estimated incoming and unassessed attempts by time window.} Unassessed attempts are concentrated after hours. In the evening and overnight periods more than nine in ten attempts remain unassessed, compared with about half during the day, and these two periods together account for about three quarters of all unassessed attempts. Because the fitted proportion of suicidal ideation is also higher in these hours, the estimated suicidal-ideation proportion is higher among unassessed attempts than among assessed calls ($10.6\%$ against $8.5\%$), and after-hours attempts make up an even larger {proportion} of the unassessed attempts involving suicidal ideation. Over the study period, this amounts to an estimated $U_{\mathcal R}=5{,}682$ unassessed attempts involving suicidal ideation ($95\%$ bootstrap interval $5{,}167$ to $6{,}163$), about $35$ per day. These are counts of attempts, and one caller may account for several of them. Within the after-hours window, the evening contributes the largest number of such attempts, whereas the overnight period has the highest proportion.

\begin{figure}[t]
\centering
\includegraphics[width=\linewidth]{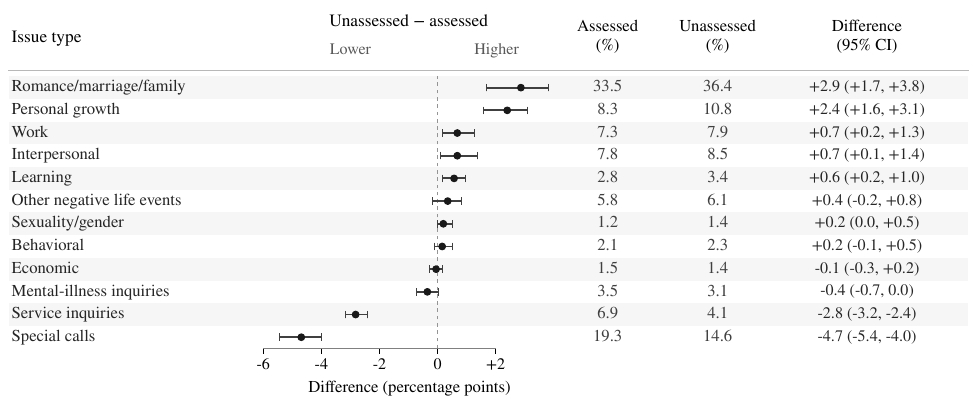}
\caption{Estimated differences in issue-type proportions between unassessed attempts and assessed calls. Points and horizontal lines show differences in percentage points and $95\%$ parametric-bootstrap intervals. Positive values indicate a higher proportion among unassessed attempts. The numeric columns report estimated proportions and differences.}
\label{fig:hidden}
\end{figure}

Figure~\ref{fig:hidden} shows the estimated differences in {issue-type proportions} between unassessed attempts and assessed calls, with $95\%$ bootstrap intervals. Unassessed attempts have a larger estimated proportion of romance, marriage, and family problems ($36.4\%$ compared with $33.5\%$) and of personal growth and other relational concerns, and a smaller proportion of special calls and service inquiries. Under the conditional independence assumption in \eqref{eq:access-model}, assessed and unassessed attempts at the same time share the same content distribution. Their estimated aggregate differences therefore reflect the timing of the attempts, most of which fall in evening and overnight hours, when the fitted proportion of relational concerns is higher and that of service inquiries is lower. By the same reasoning, unassessed attempts are weighted toward the evening caller profile of Section~\ref{sec:analysis:regimes}, namely students, callers living alone, and callers reporting loneliness or hopelessness. {In sparse periods such as the overnight hours, when few calls are assessed, the fitted issue and risk composition borrows strength across adjacent hours and days through the latent state, which also carries the temporal factors and mark information examined for \ref{q:regimes} and \ref{q:payoff}.}

\begin{table}[t]
\centering
\caption{Evening-access scenarios with incoming demand and issue composition held fixed. Each row raises the evening (17--24) access probability to the stated target. Entries are additional assessments over the $163$-day study period, with $95\%$ parametric-bootstrap intervals for the total, and the implied assessment hours per day at the mean assessed-call duration of $24.1$ minutes. Here pp denotes percentage points.}
\label{tab:counterfactual}
\small
\setlength{\tabcolsep}{4pt}
\begin{tabular}{lrcrrrr}
\toprule
& \multicolumn{4}{c}{Additional assessments ($\Delta^a$)} & \multicolumn{2}{c}{Of which} \\
\cmidrule(lr){2-5}\cmidrule(lr){6-7}
Evening access target & Total & $95\%$ interval & \shortstack[r]{Per\\day} & \shortstack[r]{Hours\\per day} & \shortstack[r]{Romance/\\family} & \shortstack[r]{Suicidal\\ideation} \\
\midrule
Current $+10$ pp & $3{,}182$ & $[3{,}143,\,3{,}215]$ & $19.5$ & $7.8$ & $1{,}201$ & $294$ \\
Current $+20$ pp & $6{,}363$ & $[6{,}287,\,6{,}430]$ & $39.0$ & $15.7$ & $2{,}403$ & $589$ \\
Daytime median ($0.48$) & $12{,}518$ & $[12{,}244,\,12{,}766]$ & $76.8$ & $30.8$ & $4{,}747$ & $1{,}170$ \\
Daytime upper quartile ($0.55$) & $14{,}776$ & $[14{,}471,\,15{,}137]$ & $90.7$ & $36.4$ & $5{,}599$ & $1{,}379$ \\
\bottomrule
\end{tabular}
\end{table}

These estimates point to evening access as a policy option, and Table~\ref{tab:counterfactual} evaluates several evening-access scenarios. Raising the evening access probability to the daytime median would yield an estimated $12{,}518$ additional assessments over the study period, about $77$ per day. Nearly two fifths of these would be romance, marriage, and family calls, and $1{,}170$ would involve suicidal ideation ($95\%$ interval $1{,}050$ to $1{,}321$). At the mean duration of an assessed call, this corresponds to about $31$ additional assessment hours per day. The scenarios hold incoming demand and issue composition fixed. If better evening access reduced repeat attempts by callers whose earlier attempts were unassessed, demand would fall and the additional assessments would be fewer than estimated.

Issue type and marks are unrecorded for unassessed attempts, so the quantities in \eqref{eq:estimands-main} are model-implied estimates and depend on the conditional independence assumption in \eqref{eq:access-model}. Under moderate departures from this assumption, the estimated proportion of suicidal ideation remains higher among unassessed attempts than among assessed calls (Section~S2.4). Risk summaries based on assessed calls alone therefore tend to understate the {proportion} of attempts with suicidal ideation.

\subsection{Latent Temporal Factors}
\label{sec:analysis:regimes}
\label{sec:analysis:trajectory}
\label{sec:analysis:loadings}

The selected model has three long-term factors and two within-day factors.
Three of them answer the second question, on temporal structure in call content (\ref{q:regimes}), and are interpreted below. Figure~\ref{fig:latent} shows the fitted factors with pointwise $95\%$ bootstrap bands. Figure~\ref{fig:signature} shows the largest issue-type and caller-characteristic loadings of the three interpreted factors. The bands of the long-term factors have mean widths of at most $0.10$ and are narrow relative to the variation in the fitted trajectories. Every loading discussed below has a $95\%$ bootstrap interval that excludes zero. Figures~S3 to~S5 of the Supplementary Material report the full sets of loadings for all five factors.

\begin{figure}[t]
\centering
\begin{subfigure}[t]{0.57\linewidth}
\centering
\includegraphics[width=\linewidth]{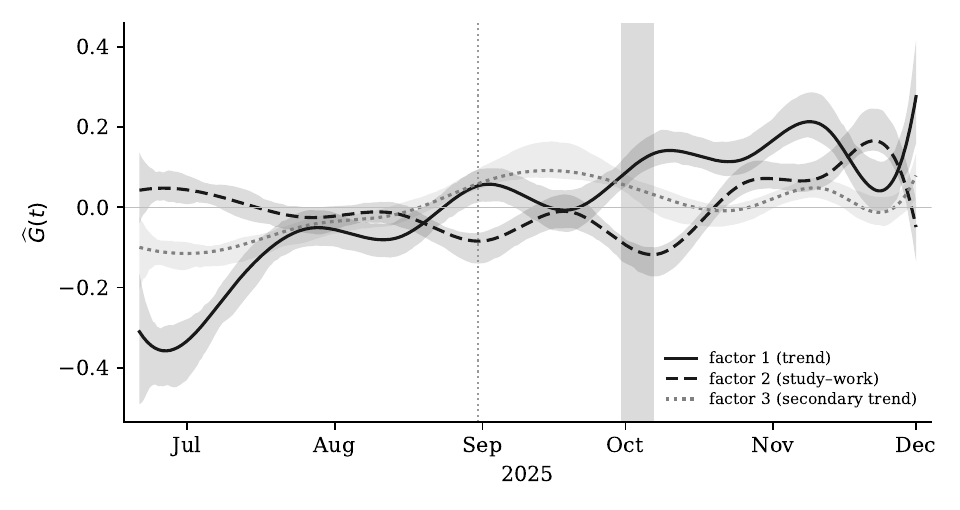}
\caption{Long-term factors $\widehat{\mathbf G}(t)$.}
\label{fig:trajectory}
\end{subfigure}\hfill
\begin{subfigure}[t]{0.41\linewidth}
\centering
\includegraphics[width=\linewidth]{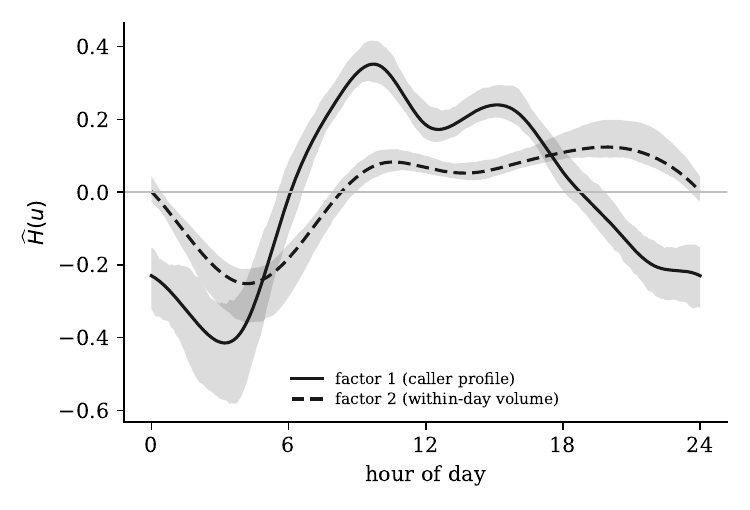}
\caption{Within-day factors $\widehat{\mathbf H}(u)$.}
\label{fig:H}
\end{subfigure}
\caption{The fitted two-scale latent state with pointwise $95\%$ bootstrap bands. Panel (a) shows the long-term factors (factor~1 solid, factor~2 dashed, factor~3 dotted), with the public-holiday period shaded and a dotted vertical line marking the start of the autumn semester. Panel (b) shows the within-day factors (first solid, second dashed).}
\label{fig:latent}
\end{figure}

\begin{figure}[t]
\centering
\includegraphics[width=\linewidth]{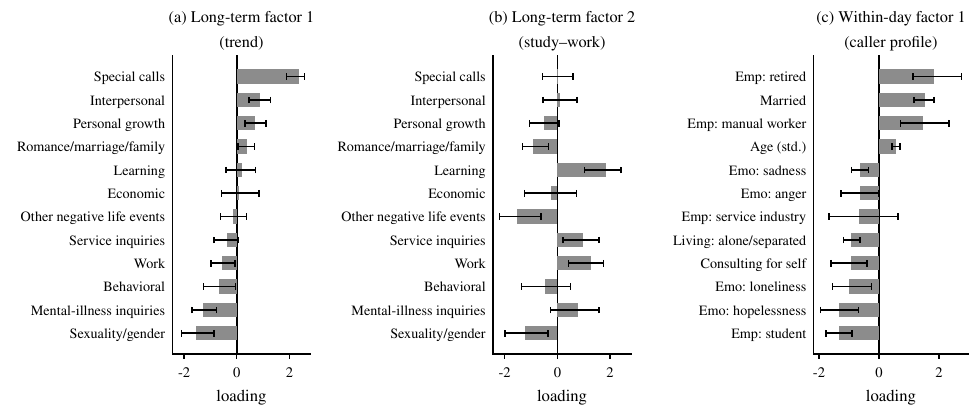}
\caption{Loadings of the three interpreted latent temporal factors. Panels (a) and (b) show the twelve issue-type loadings on long-term factors~1 and~2 in a common row order. Panel (c) shows a selected subset of mark loadings on the caller-profile factor, with positive values corresponding to daytime hours and negative values to evening hours. All loadings are shown with $95\%$ parametric-bootstrap intervals.}
\label{fig:signature}
\end{figure}

The first long-term factor (solid curve in Figure~\ref{fig:latent}a) is termed the {trend factor}. It rises steadily from about $-0.36$ in late June to about $+0.28$ in late November. Sexuality and gender problems, mental-illness inquiries, and behavioral and work problems display large negative loadings. Interpersonal problems, personal growth issues, and special calls such as silent or prank contacts exhibit large positive loadings (Figure~\ref{fig:signature}a), meaning their fitted proportions grow over the study period. {Refitting the model without the special-calls type leaves the long-term factors virtually unchanged, yielding aligned correlations between $0.92$ and $0.99$ (Table~S5 of the Supplementary Material).} This factor therefore captures long-term {change in the issue composition of incoming attempts}. Distress severity is examined separately in Section~\ref{sec:analysis:crisis} through the suicidal-ideation proportion.

The second long-term factor (dashed curve in Figure~\ref{fig:latent}a) is termed the study--work factor and has about half that amplitude. It rises around mid-September, when the autumn semester begins. It then falls to its minimum during the major public-holiday period at the beginning of October and recovers through November. Learning, work, and service-inquiry calls have positive loadings. Other negative life events, sexuality and gender problems, and romance, marriage, and family problems have negative loadings (Figure~\ref{fig:signature}b). The factor therefore describes a rise in the proportions of learning, work, and service-inquiry calls at the start of the semester and a fall during the holiday.

The first within-day factor (solid curve in Figure~\ref{fig:latent}b) is termed the caller-profile factor. It rises to a late-morning maximum and declines through the evening, so positive values correspond to daytime hours. Marriage, retirement, and manual employment have positive loadings (Figure~\ref{fig:signature}c), as do service inquiries and special calls. The factor's hour-of-day profile closely tracks the hourly proportion of service-inquiry calls (correlation $0.91$). Student status, living alone, calling on one's own behalf, and reported loneliness and hopelessness have negative loadings. These characteristics are therefore more common in the evening, when the assessment proportion is lowest. These differences between evening and daytime callers underlie the content differences between unassessed attempts and assessed calls reported in Section~\ref{sec:analysis:hidden}.

The two remaining factors capture additional temporal variation. The second within-day factor (dashed curve in Figure~\ref{fig:latent}b) carries the largest within-day demand loading. Its hour-of-day profile tracks the observed hourly attempt counts (correlation $0.79$), so it mainly describes within-day variation in demand. The third long-term factor (dotted curve in Figure~\ref{fig:latent}a) is a secondary trend of smaller amplitude in issue composition.

\subsection{Information Gain and Robustness}
\label{sec:analysis:borrowing}

The third question concerns the information contributed by the marks (\ref{q:payoff}). The marks improve the precision of the fitted latent state {without degrading the held-out log likelihood of the issue labels}. The precision gains are consistent with the information decomposition in Section~\ref{sec:ident:info}. We compare the joint estimator with the events-only estimator on matched bootstrap datasets. The mean reduction in pointwise $95\%$ band width is $18\%$, $49\%$, and $38\%$ for the three long-term factors and $29\%$ and $20\%$ for the two within-day factors (Figure~\ref{fig:borrowing-main}). 
The largest reduction in uncertainty occurs for the study--work factor, where issue-labeled event data are relatively sparse. Across all factors, the uncertainty reduction gained from incorporating marks is most pronounced in the quartile with the fewest assessed calls. Furthermore, refitting the model after omitting each mark variable in turn shows that no single mark drives the gains in held-out fit (Section~S2.5 of the Supplementary Material).

\begin{figure}[t]
\centering
\includegraphics[width=\linewidth]{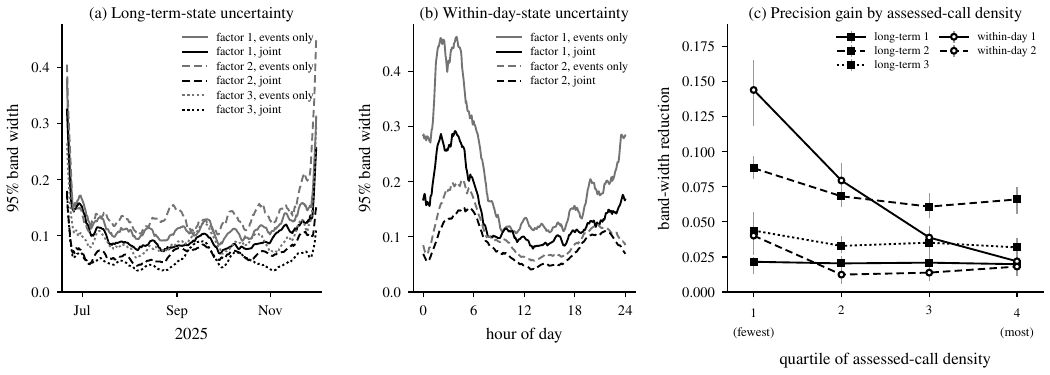}
\caption{Borrowing strength from the marks. Panels (a) and (b) show pointwise $95\%$ bootstrap band widths of the long-term and within-day factors under the events-only and joint estimators on matched bootstrap datasets. Panel (c) shows the mean reduction in band width, events only minus joint, by quartile of assessed-call density for each factor, with $95\%$ resampling intervals.}
\label{fig:borrowing-main}
\end{figure}

\label{sec:analysis:crisis}
\begin{figure}[tp]
\centering
\includegraphics[width=\linewidth]{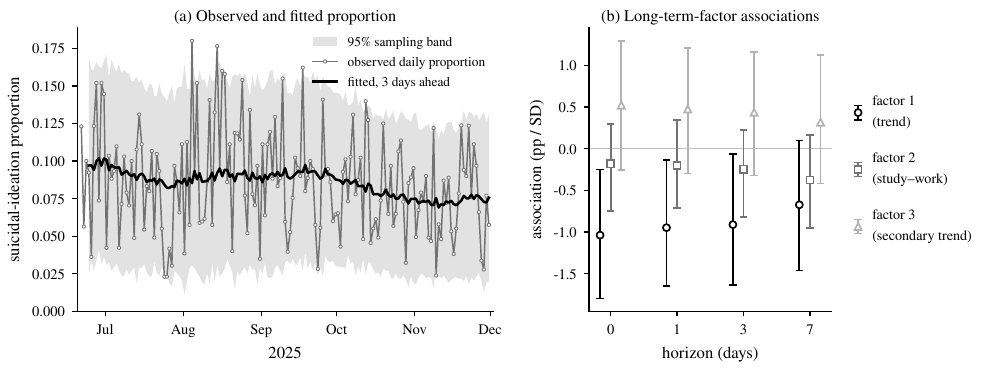}
\caption{Retrospective summary of the suicidal-ideation proportion among assessed calls. Panel (a) shows the daily observed proportion, the fitted three-day-ahead proportion, and a pointwise $95\%$ binomial sampling band. Panel (b) shows the association of the three long-term factors with this proportion at horizons of $0$, $1$, $3$, and $7$ days, in percentage points per standard deviation, with $95\%$ moving-block bootstrap intervals.}
\label{fig:crisis-main}
\end{figure}

The {trend factor} also tracks the proportion of assessed calls with recorded suicidal ideation. The mean of this daily proportion is $8.6\%$. It was highest early in the study period and declined as the issue composition changed. With about $87$ issue-labeled calls per day, its day-to-day variation includes substantial binomial sampling noise. We therefore relate the daily proportion to current and lagged long-term factors through binomial regressions. A one-standard-deviation increase in the {trend factor} is associated with a decrease of $1.04$ percentage points in the same-day proportion ($95\%$ interval $0.25$ to $1.79$). The association persists at lags of one and three days and weakens by seven days. The bootstrap intervals for the other two long-term factors include zero (Figure~\ref{fig:crisis-main}). A refit without the crisis-severity mark reproduces both the long-term state and the association. When the long-term state is estimated from past data only, short-term forecasts of the daily proportion are essentially unchanged (Section~S2.6 of the Supplementary Material). The state is therefore useful for reviewing past changes in caller risk. Early warning would require updating it as new data arrive.

\label{sec:analysis:checks}
\label{sec:analysis:fit}
\label{sec:analysis:sensitivity}
\label{sec:analysis:repro}
Diagnostics and sensitivity analyses support the reported structure. We evaluate model adequacy through two goodness-of-fit checks against the observed records. First, aggregating the fitted issue-type probabilities by week closely reproduces the observed weekly proportions across all twelve categories, with differences never exceeding $5$ percentage points. Second, the randomized probability integral transform of the fitted conditional distribution for each discrete mark is approximately uniform, indicating adequate calibration. In addition, refitting the model under alternative coding choices and model specifications yields long-term and within-day trajectories that maintain aligned correlations of at least $0.92$ with the reported fit. Sections~S2.3 to~S2.6 of the Supplementary Material provide additional fit diagnostics, sensitivity analyses, and predictive evaluations.

\section{Discussion}\label{sec:discussion}


In this article, we developed the JM-DFM, a joint dynamic factor model for marked point processes in which event times are fully recorded but marks are only partially observed because of capacity constraints. A shared low-dimensional latent state drives the attempt intensity, the issue composition, and the mark distributions, and the observed-data likelihood integrates out the unrecorded content of unassessed attempts under an ignorable assessment mechanism. The theoretical results in Section~\ref{sec:theory} establish identifiability and consistency of the sieve estimator. In the application to the 12356 hotline, attempts are concentrated in evening and overnight hours, when assessment probabilities are lowest (\ref{q:unmet}). Under the observation assumptions, the estimated proportions of relational concerns and suicidal ideation are higher among unassessed attempts than among assessed calls. The fitted latent factors (\ref{q:regimes}) summarize long-term and within-day variation in issue composition and caller characteristics, including a {trend factor}, a study--work factor, and a caller-profile factor whose daytime-to-evening shift underlies the content differences between assessed and unassessed calls. Jointly modeling the marks with the attempt process (\ref{q:payoff}) narrows the uncertainty bands of the latent state, most during periods with few issue-labeled calls.


Two methodological limitations suggest directions for future research. First, the present formulation uses a moderate number of structured marks, $P=10$ in our application, whereas hotlines often store call audio and text transcripts. One natural extension is to extract dense representations, such as text embeddings or acoustic summaries, and include them as additional marks in \eqref{eq:mark-model}. Analyzing such data within our framework would require high-dimensional factor asymptotics in which $P$ grows with the sample size, together with regularized estimation of the loading matrices. Alternatively, one could treat the audio or transcript directly as an unstructured mark and specify a flexible conditional density. That direction would raise new questions of nonparametric identifiability and of convergence rates for the estimated latent trajectories.

Second, the sieve estimator is fitted to a fixed observation window, so the estimated latent state is a retrospective summary. Such a fit is suited to historical evaluation and capacity planning, whereas real-time monitoring requires sequential updating. Stochastic-approximation or recursive filtering methods could update the estimated latent state as each day's attempts and assessment records arrive. Prospective monitoring would then require establishing the stability of the recursive updates and evaluating their performance against prespecified operational targets (Section~\ref{sec:analysis:crisis}).

The observation structure treated in this paper is common to public and clinical systems in which requests are logged automatically but detailed triage information depends on staff availability. Examples include emergency dispatch centers, digital telehealth platforms, and centralized mental health intake queues. The joint likelihood and estimation strategy developed here apply to such systems when the request stream is fully observed and access to assessment is independent of the unrecorded content given the observed operational variables.

\paragraph{Supplementary Materials.}
The online supplement contains the simulation studies (Section~S1),
additional results for the hotline analysis (Section~S2), and proofs of the
theoretical results (Section~S3). The call-level records from the 12356
hotline used in this study are confidential. Code reproducing the simulation
studies is available at\par\nopagebreak
{\centering\url{https://anonymous.4open.science/r/jmdfm-replication-12A0/}\par}
\noindent A synthetic dataset and code illustrating the analysis will be
added to the same repository.

\paragraph{Acknowledgments.}
Zhang is supported by the National Natural Science Foundation of China (Grant No.~12671345). Ouyang is partially supported by the Hong Kong Early Career Scheme (ECS) Grant \#27308125. The authors thank the National Mental Health Center of China, a collaborating institution on the funded project, for providing the de-identified 12356 hotline records under a collaboration agreement with East China Normal University. As a secondary analysis of anonymous records, the study qualified for exemption from ethics review under Article~32 of China's 2023 Measures for Ethical Review of Life Sciences and Medical Research Involving Humans.

\paragraph{Declaration of Generative AI Use.}
The authors used ChatGPT 6 Astra and Claude Pro to assist with language editing of the manuscript and with the code for the numerical studies, in order to improve the clarity of the writing and the efficiency of the implementation. The authors carefully reviewed and verified the resulting text and code, and they take full responsibility for the contents of this article.

\paragraph{Disclosure Statement.}
The authors report there are no competing interests to declare.

\setstretch{.85}
\bibliographystyle{apalike}
\bibliography{refs}

@misc{chinadaily2025hotline,
  author       = {{China Daily}},
  title        = {{National Health Commission} to fully implement eight initiatives this year},
  year         = {2025},
  howpublished = {\url{https://english.www.gov.cn/news/202502/14/content_WS67aeea23c6d0868f4e8efa28.html}},
  note         = {Published February 14, 2025. Accessed September 15, 2026}
}

@article{chen2025dynamic,
  title={Dynamic Factor Analysis of High-Dimensional Recurrent Events},
  author={Chen, Fangyi and Chen, Yunxiao and Ying, Zhiliang and Zhou, Kangjie},
  journal={Biometrika},
  volume={112},
  number={3},
  pages={asaf028},
  year={2025},
  doi={10.1093/biomet/asaf028}
}

@article{lin2000semiparametric,
  title={Semiparametric regression for the mean and rate functions of recurrent events},
  author={Lin, Danyu Y and Wei, Lee-Jen and Yang, I and Ying, Zhiliang},
  journal={Journal of the Royal Statistical Society: Series B (Statistical Methodology)},
  volume={62},
  number={4},
  pages={711--730},
  year={2000},
  publisher={Wiley Online Library}
}

@article{zhang2022joint,
  title={Joint latent space models for network data with high-dimensional node variables},
  author={Zhang, Xuefei and Xu, Gongjun and Zhu, Ji},
  journal={Biometrika},
  volume={109},
  number={3},
  pages={707--720},
  year={2022},
  publisher={Oxford University Press}
}

@article{chen2025dynamicfactor,
  title={A Dynamic Factor Model for Multivariate Counting Process Data},
  author={Chen, Fangyi and Ling, Hok Kan and Ying, Zhiliang},
  journal={arXiv preprint arXiv:2503.01081},
  year={2025}
}

@article{wang2022maximum,
title = {Maximum likelihood estimation and inference for high dimensional generalized factor models with application to factor-augmented regressions},
journal = {Journal of Econometrics},
volume = {229},
number = {1},
pages = {180-200},
year = {2022},
issn = {0304-4076},
author = {Fa Wang},
}

@article{ogata1988statistical,
  author  = {Ogata, Yosihiko},
  title   = {Statistical Models for Earthquake Occurrences and Residual Analysis for Point Processes},
  journal = {Journal of the American Statistical Association},
  year    = {1988},
  volume  = {83},
  number  = {401},
  pages   = {9--27},
  doi     = {10.1080/01621459.1988.10478560}
}

@inproceedings{du2016recurrent,
  author    = {Du, Nan and Dai, Hanjun and Trivedi, Rakshit and Upadhyay, Utkarsh and Gomez-Rodriguez, Manuel and Song, Le},
  title     = {Recurrent Marked Temporal Point Processes: Embedding Event History to Vector},
  booktitle = {Proceedings of the 22nd ACM SIGKDD International Conference on Knowledge Discovery and Data Mining},
  pages     = {1555--1564},
  year      = {2016},
  doi       = {10.1145/2939672.2939875}
}

@article{andersen1982cox,
  author  = {Andersen, Per Kragh and Gill, Richard D.},
  title   = {Cox's Regression Model for Counting Processes: A Large Sample Study},
  journal = {The Annals of Statistics},
  year    = {1982},
  volume  = {10},
  number  = {4},
  pages   = {1100--1120}
}

@book{coxisham1980point,
  author    = {Cox, David R. and Isham, Valerie},
  title     = {Point Processes},
  year      = {1980},
  publisher = {Chapman \& Hall},
  address   = {London}
}

@book{cook2007statistical,
  author    = {Cook, Richard J. and Lawless, Jerald F.},
  title     = {The Statistical Analysis of Recurrent Events},
  year      = {2007},
  publisher = {Springer},
  address   = {New York}
}

@book{daley2003introduction,
  author    = {Daley, Daryl J. and Vere-Jones, David},
  title     = {An Introduction to the Theory of Point Processes, Volume I: Elementary Theory and Methods},
  edition   = {2nd},
  year      = {2003},
  publisher = {Springer},
  address   = {New York}
}

@article{lin2001semiparametric,
  author  = {Lin, Danyu Y. and Wei, Lee-Jen and Ying, Zhiliang},
  title   = {Semiparametric Transformation Models for Point Processes},
  journal = {Journal of the American Statistical Association},
  year    = {2001},
  volume  = {96},
  number  = {454},
  pages   = {620--628}
}

@article{wulfsohn1997joint,
  author  = {Wulfsohn, Michael S. and Tsiatis, Anastasios A.},
  title   = {A Joint Model for Survival and Longitudinal Data Measured with Error},
  journal = {Biometrics},
  year    = {1997},
  volume  = {53},
  number  = {1},
  pages   = {330--339}
}

@article{henderson2000joint,
  author  = {Henderson, Robin and Diggle, Peter and Dobson, Annette},
  title   = {Joint Modelling of Longitudinal Measurements and Event Time Data},
  journal = {Biostatistics},
  year    = {2000},
  volume  = {1},
  number  = {4},
  pages   = {465--480}
}

@article{cai2010semiparametric,
  author  = {Cai, Jianwen and Zeng, Donglin and Pan, Wenqin},
  title   = {Semiparametric Proportional Means Model for Marker Data Contingent on Recurrent Event},
  journal = {Lifetime Data Analysis},
  year    = {2010},
  volume  = {16},
  number  = {2},
  pages   = {250--270}
}

@article{liu2004shared,
  author  = {Liu, Lei and Wolfe, Robert A. and Huang, Xuelin},
  title   = {Shared Frailty Models for Recurrent Events and a Terminal Event},
  journal = {Biometrics},
  year    = {2004},
  volume  = {60},
  number  = {3},
  pages   = {747--756}
}

@article{kim2012joint,
  author  = {Kim, Sehee and Zeng, Donglin and Chambless, Lloyd and Li, Yi},
  title   = {Joint Models of Longitudinal Data and Recurrent Events with Informative Terminal Event},
  journal = {Statistics in Biosciences},
  year    = {2012},
  volume  = {4},
  number  = {2},
  pages   = {262--281}
}

@article{hawkes1971spectra,
  author  = {Hawkes, Alan G.},
  title   = {Spectra of Some Self-Exciting and Mutually Exciting Point Processes},
  journal = {Biometrika},
  year    = {1971},
  volume  = {58},
  number  = {1},
  pages   = {83--90}
}

@article{stock2002forecasting,
  author  = {Stock, James H. and Watson, Mark W.},
  title   = {Forecasting Using Principal Components from a Large Number of Predictors},
  journal = {Journal of the American Statistical Association},
  year    = {2002},
  volume  = {97},
  number  = {460},
  pages   = {1167--1179}
}

@article{kaiser1958varimax,
  author  = {Kaiser, Henry F.},
  title   = {The Varimax Criterion for Analytic Rotation in Factor Analysis},
  journal = {Psychometrika},
  year    = {1958},
  volume  = {23},
  number  = {3},
  pages   = {187--200}
}

@article{bai2003inferential,
  author  = {Bai, Jushan},
  title   = {Inferential Theory for Factor Models of Large Dimensions},
  journal = {Econometrica},
  year    = {2003},
  volume  = {71},
  number  = {1},
  pages   = {135--171}
}

@article{bai2012statistical,
  author  = {Bai, Jushan and Li, Kunpeng},
  title   = {Statistical Analysis of Factor Models of High Dimension},
  journal = {The Annals of Statistics},
  year    = {2012},
  volume  = {40},
  number  = {1},
  pages   = {436--465},
  doi     = {10.1214/11-AOS966}
}

@article{forni2000generalized,
  author  = {Forni, Mario and Hallin, Marc and Lippi, Marco and Reichlin, Lucrezia},
  title   = {The Generalized Dynamic-Factor Model: Identification and Estimation},
  journal = {Review of Economics and Statistics},
  year    = {2000},
  volume  = {82},
  number  = {4},
  pages   = {540--554}
}

@article{chenli2022determining,
  author  = {Chen, Yunxiao and Li, Xiaoou},
  title   = {Determining the Number of Factors in High-Dimensional Generalized Latent Factor Models},
  journal = {Biometrika},
  year    = {2022},
  volume  = {109},
  number  = {3},
  pages   = {769--782}
}

@book{bartholomew2011latent,
  author    = {Bartholomew, David J. and Knott, Martin and Moustaki, Irini},
  title     = {Latent Variable Models and Factor Analysis: A Unified Approach},
  edition   = {3rd},
  year      = {2011},
  publisher = {Wiley},
  address   = {Chichester}
}

@inproceedings{shchur2021neural,
  author    = {Shchur, Oleksandr and T{\"u}rkmen, Ali Caner and Januschowski, Tim and G{\"u}nnemann, Stephan},
  title     = {Neural Temporal Point Processes: A Review},
  booktitle = {Proceedings of the 30th International Joint Conference on Artificial Intelligence},
  pages     = {4585--4593},
  year      = {2021}
}

@article{tadmon2023differential,
  author  = {Tadmon, Daniel and Bearman, Peter S.},
  title   = {Differential Spatial-Social Accessibility to Mental Health Care and Suicide},
  journal = {Proceedings of the National Academy of Sciences},
  year    = {2023},
  volume  = {120},
  number  = {19},
  pages   = {e2301304120},
  doi     = {10.1073/pnas.2301304120}
}

@article{brulhart2021mental,
  author  = {Br{\"u}lhart, Marius and Klotzb{\"u}cher, Valentin and Lalive, Rafael and Reich, Stephanie K.},
  title   = {Mental Health Concerns during the {COVID-19} Pandemic as Revealed by Helpline Calls},
  journal = {Nature},
  year    = {2021},
  volume  = {600},
  number  = {7887},
  pages   = {121--126},
  doi     = {10.1038/s41586-021-04099-6}
}

@article{ehrlich2023trends,
  author  = {Kandula, Sasikiran and Higgins, Johnathan and Goldstein, Alena and Gould, Madelyn S. and Olfson, Mark and Keyes, Katherine M. and Shaman, Jeffrey},
  title   = {Trends in Crisis Hotline Call Rates and Suicide Mortality in the {United States}},
  journal = {Psychiatric Services},
  year    = {2023},
  volume  = {74},
  number  = {9},
  pages   = {978--981},
  doi     = {10.1176/appi.ps.20220199}
}

@article{marco2017spatio,
  author  = {Marco, Miriam and L{\'o}pez-Qu{\'\i}lez, Antonio and Conesa, David and Gracia, Enrique and Lila, Marisol},
  title   = {Spatio-Temporal Analysis of Suicide-Related Emergency Calls},
  journal = {International Journal of Environmental Research and Public Health},
  year    = {2017},
  volume  = {14},
  number  = {7},
  pages   = {735}
}

@article{rubin1976inference,
  author  = {Rubin, Donald B.},
  title   = {Inference and Missing Data},
  journal = {Biometrika},
  year    = {1976},
  volume  = {63},
  number  = {3},
  pages   = {581--592},
  doi     = {10.1093/biomet/63.3.581}
}

@article{heitjan1991ignorability,
  author  = {Heitjan, Daniel F. and Rubin, Donald B.},
  title   = {Ignorability and Coarse Data},
  journal = {The Annals of Statistics},
  year    = {1991},
  volume  = {19},
  number  = {4},
  pages   = {2244--2253},
  doi     = {10.1214/aos/1176348396}
}

@article{weinberg2007bayesian,
  author  = {Weinberg, Jonathan and Brown, Lawrence D. and Stroud, Jonathan R.},
  title   = {Bayesian Forecasting of an Inhomogeneous {P}oisson Process with Applications to Call Center Data},
  journal = {Journal of the American Statistical Association},
  year    = {2007},
  volume  = {102},
  number  = {480},
  pages   = {1185--1198},
  doi     = {10.1198/016214506000001455}
}

@article{matteson2011forecasting,
  author  = {Matteson, David S. and McLean, Mathew W. and Woodard, Dawn B. and Henderson, Shane G.},
  title   = {Forecasting Emergency Medical Service Call Arrival Rates},
  journal = {The Annals of Applied Statistics},
  year    = {2011},
  volume  = {5},
  number  = {2B},
  pages   = {1379--1406},
  doi     = {10.1214/10-AOAS442}
}

@article{zhou2015spatio,
  author  = {Zhou, Zhengyi and Matteson, David S. and Woodard, Dawn B. and Henderson, Shane G. and Micheas, Athanasios C.},
  title   = {A Spatio-Temporal Point Process Model for Ambulance Demand},
  journal = {Journal of the American Statistical Association},
  year    = {2015},
  volume  = {110},
  number  = {509},
  pages   = {6--15},
  doi     = {10.1080/01621459.2014.941466}
}

@article{mohler2011self,
  author  = {Mohler, George O. and Short, Martin B. and Brantingham, P. Jeffrey and Schoenberg, Frederic Paik and Tita, George E.},
  title   = {Self-Exciting Point Process Modeling of Crime},
  journal = {Journal of the American Statistical Association},
  year    = {2011},
  volume  = {106},
  number  = {493},
  pages   = {100--108},
  doi     = {10.1198/jasa.2011.ap09546}
}

@article{reinhart2018review,
  author  = {Reinhart, Alex},
  title   = {A Review of Self-Exciting Spatio-Temporal Point Processes and Their Applications},
  journal = {Statistical Science},
  year    = {2018},
  volume  = {33},
  number  = {3},
  pages   = {299--318}
}

@book{efron1993introduction,
  title={An Introduction to the Bootstrap},
  author={Efron, Bradley and Tibshirani, Robert J.},
  year={1993},
  publisher={Chapman \& Hall},
  address={New York}
}

@book{davison1997bootstrap,
  title={Bootstrap Methods and Their Application},
  author={Davison, Anthony C. and Hinkley, David V.},
  year={1997},
  publisher={Cambridge University Press},
  address={Cambridge}
}

@article{stone1982optimal,
  title={Optimal global rates of convergence for nonparametric regression},
  author={Stone, Charles J},
  journal={The Annals of Statistics},
  volume={10},
  number={4},
  pages={1040--1053},
  year={1982}
}

@article{hausman1978specification,
  author  = {Hausman, J. A.},
  title   = {Specification Tests in Econometrics},
  journal = {Econometrica},
  year    = {1978},
  volume  = {46},
  number  = {6},
  pages   = {1251--1271},
  doi     = {10.2307/1913827}
}

@article{green1991james,
  author  = {Green, Edwin J. and Strawderman, William E.},
  title   = {A {J}ames--{S}tein type estimator for combining unbiased and possibly biased estimators},
  journal = {Journal of the American Statistical Association},
  year    = {1991},
  volume  = {86},
  number  = {416},
  pages   = {1001--1006},
  doi     = {10.1080/01621459.1991.10475144}
}

\end{document}